\documentclass[preprint,12pt]{article}

\usepackage{amsmath,amssymb,mathtools}
\usepackage{siunitx}
\usepackage{graphicx}
\usepackage{subcaption} 
\usepackage{tikz}         
\usepackage{float}         
\usepackage{booktabs}
\usepackage{hyperref}
\usepackage[margin=1in]{geometry}
\usepackage{xcolor}
\usepackage[utf8]{inputenc}
\usepackage{enumitem}  
\usepackage{authblk}
\DeclareSIUnit{\photon}{photon}	

\title{From Light Diffusion to Photocatalytic Rates: Compact Scaling Laws for Strongly Scattering Porous Slabs}
	
\author[1,*]{Renaud A. L. Vall\'{e}e}
\author[1]{R\'{e}nal Backov}
\affil[1]{Univ. Bordeaux, CNRS, CRPP, UMR5031, 33600 Pessac, France}
\affil[*]{Corresponding author: renaud.vallee@u-bordeaux.fr}
	
\begin{document}
\maketitle
		
\begin{abstract}
		
		Light transport in strongly scattering porous photocatalytic materials governs
		the spatial distribution of absorbed photons and therefore the generation of
		charge carriers driving photocatalytic reactions. Yet translating measured
		optical properties of such media into intrinsic reaction rate constants remains
		challenging, as it requires simultaneously accounting for multiple scattering,
		boundary losses, photochemical efficiency, and surface kinetics. Here we develop
		a compact analytical framework that unifies these processes for
		nanoparticle-loaded photocatalytic slabs. Using a finite-slab diffusion model
		with extrapolated boundaries, we derive closed-form expressions for the fluence
		field and couple them to a photochemical quantum efficiency and first-order
		surface kinetics. The resulting predictors yield intrinsic volumetric and areal
		rate constants whose dependence on the transport mean free path, optical
		thickness, and surface-to-volume ratio emerges transparently.
		
		Validation against Monte-Carlo photon-migration simulations shows that the
		diffusion approximation reproduces the fluence and generation-rate profiles with
		a modest multiplicative mismatch, typically within a factor of $\sim$1.20–1.39,
		depending on the anisotropy and scattering phase function. This level of
		agreement is consistent with the known limits of the diffusion approximation and
		is sufficient to enable reliable, design-oriented predictions.
		
		The analytical descriptors introduced here—such as $\ell^{*}$, $S/V$, and the
		extrapolation length $z_b$—are general, physically interpretable, and directly
		integrable into data-driven optimisation and geometry-engineered reactor design.
		The framework thus provides a versatile and physically grounded tool for
		photocatalytic systems across diverse applications, including VOC
		photo-degradation, indoor air purification, and solar-fuel production.
	
\end{abstract}
				
		
	\section{Introduction}
	
	Porous monolithic photocatalytic slabs have emerged as a versatile platform for
	solar-driven synthesis, indoor air purification, and VOC photo-degradation,
	offering high specific surface area, structural robustness, and scalable reactor potential
	\cite{Vardon2025,Layan2025}. In these materials, incident photons undergo intense
	multiple scattering: the optical properties of the medium (absorption coefficient $\mu_a$,
	reduced scattering coefficient $\mu_s'$, and anisotropy factor $g$) critically determine
	the spatial distribution of photon absorption and thus the generation of charge carriers.
	Diffuse-reflectance and transmittance spectroscopy routinely yield $\mu_a$ and $\mu_s'$
	\cite{Li2024,Grabtchak2012,Spinelli2007,Martelli2007}, but a quantitative mapping from
	these optical parameters to intrinsic catalytic rate constants remains elusive.
	
	Current modeling strategies fall into two broad classes. On one hand, full
	radiative-transfer simulations—such as Monte-Carlo or discrete-ordinates methods—can
	accurately predict fluence fields in scattering media, even for anisotropic scattering
	or optically thin slabs \cite{Wang1995,Kienle1997}. However these methods are computationally
	demanding, making them impractical for extensive parameter sweeps, reactor optimisation,
	or data-driven design. On the other hand, simple Beer–Lambert–type attenuation models,
	often employed in photocatalysis, neglect scattering and boundary effects altogether,
	and thus fail when applied to thick or strongly scattering composites such as aerogels
	or monoliths \cite{deBoer2019}.
	
	The Diffusion approximation (DA) offers an attractive
	intermediate path. When $\mu_s' \gg \mu_a$ and the transport mean free path
	$\ell^{*} = 1 / \mu_s'$ is much smaller than the slab thickness, photon propagation can be
	described by a diffusion equation \cite{CaseZweifel1967,Ishimaru1978}. Classical slab 
	solutions with extrapolated (partial-current) boundary conditions are well established
	\cite{Haskell1994,Paasschens2008}, and they reproduce Monte-Carlo fluence fields under many
	scattering conditions. Nonetheless, these optical solutions have rarely been
	integrated with photocatalytic reaction kinetics.
	
	In parallel, the kinetics of heterogeneous photocatalysis has traditionally been described
	using the Langmuir–Hinshelwood model (L–H) or pseudo-first-order
	rate laws. Many studies of liquid-phase or gas-phase pollutant degradation fit time-series
	data with such empirical models \cite{Tran2023,Sun2018,CameraRoda2025}. However, as critically
	discussed by Ollis and co-workers \cite{Ollis2018}, L–H fits rarely reveal mechanistic
	insight, because the same mathematical form can emerge from different underlying mechanisms,
	and the parameters (adsorption constant, catalytic rate) may depend strongly on light
	intensity, oxygen concentration, mass transfer limitations, or intermediates. More refined
	kinetic analyses emphasize that neglecting spatial inhomogeneity of light intensity or
	neglecting mass transfer often leads to misinterpretation of observed kinetics
	\cite{Bloh2019}. As a result, existing kinetic models generally assume uniform illumination
	or treat photon absorption implicitly, which limits their predictive power for structured or
	strongly scattering photocatalytic media.
	
	A recent experimental study highlights the importance of coupling optical scattering and
	catalysis. Layan \emph{et al.} \cite{Layan2025} fabricated self-standing TiO$_2$--SiO$_2$
	monoliths with multiscale porosity and demonstrated efficient VOC mineralisation
	($\approx 80 \, \%$ propan-2-one conversion) at very low Ti content (Ti/Si ratio 0.08--0.15),
	suggesting that the bulk scattering properties—rather than merely the surface area—
	dominate the photocatalytic response. This underscores the need for a model that reliably
	maps optical–transport parameters into catalytic rates.
	
	Moreover, emerging data-driven and geometry-engineering approaches reinforce this need.
	Machine-learning studies have shown that including physics-based descriptors as input features
	improves the predictive power of photocatalytic activity models \cite{Liu2023}, while
	topology-optimised 3D-printed reactors highlight the role of internal geometry in controlling
	optical path-length distributions \cite{Brunser2023}. However both strategies currently rely on
	either large experimental datasets or expensive simulations, unless supported by compact,
	physically meaningful descriptors.
	
	\medskip
	\noindent  
	To address this gap, we derive an analytical framework that combines photon diffusion,
	photochemical efficiency, and surface reaction kinetics in a unified, self-consistent way.
	Starting from the finite-slab Green's function for an isotropic internal source with
	extrapolated boundary conditions, we introduce an internal quantum efficiency
	$\phi_{\mathrm{int}}$ that converts absorbed photons into reactive charge carriers.
	By coupling this to site-limited or Langmuir–Hinshelwood–type kinetics, we obtain
	closed-form expressions for compact intrinsic volumetric
	(Eqs.~\ref{eq:kVcompact}) and areal (Eqs.~\ref{eq:kAcompact}) rate constants. These expressions provide a direct,
	quantitative mapping between bulk optical parameters ($\mu_a, \mu_s', g$) and catalytic
	performance.
	
	We demonstrate the framework using the TiO$_2$--SiO$_2$ aerogels from \cite{Layan2025},
	and show it is readily extensible to other porous photocatalytic media such as
	polymer-bound metal oxides \cite{Liu2023} and 3D-printed reactors \cite{Brunser2023}. The
	compact predictors enable efficient parameter exploration, uncertainty analysis, and
	design-oriented optimisation, making them a versatile tool for next-generation photocatalytic
	devices across applications (VOC degradation, water purification, solar-fuel production).
	
	\medskip
	\noindent The manuscript is organised as follows: Section 2 presents the diffusion theory
	and Green’s function derivation; Section 3 introduces the optical-to-kinetic mapping; Section 4
	describes the extraction of optical properties (EMA, Mie calculations); Section 5 benchmarks
	the diffusion solution against Monte-Carlo simulations and discusses the limits of the
	approximation (anisotropy, dependent scattering, boundary conditions); Section 6 reports
	intrinsic rate constants and offers design guidelines; finally Section 7 summarises
	conclusions and outlines future work.

	\section{Theory: steady‑state diffusion with extrapolated boundaries}
	\label{sec:theory}
	The propagation of photons in a highly scattering porous slab can be described, in the regime $\mu_s'\gg\mu_a$, by the steady‑state diffusion equation. Throughout this section we retain photon‑unit conventions, i.e. the fluence rate $\Phi(z,\lambda)$ carries the dimension \si{\photon\per\square\meter\per\second}.
	
	\subsection{Diffusion equation and material parameters}
	For a monochromatic field of wavelength $\lambda$ the diffusion approximation reads
	\begin{equation}
		- D(\lambda)\,\nabla^{2}\Phi(z,\lambda)+\mu_{a}(\lambda)\,\Phi(z,\lambda)=S(\mathbf{r},\lambda),
		\label{eq:diffusion}
	\end{equation}
	where the diffusion coefficient is
	\begin{equation}
		D(\lambda)=\frac{1}{3\big[\mu_{a}(\lambda)+\mu_{s}'(\lambda)\big]},
	\end{equation}
	$\mu_{a}$ and $\mu_{s}'$ are the absorption and reduced‑scattering coefficients (units \si{\per\meter}), and the source term $S$ has units \si{\photon\per\cubic\meter\per\second}. The transport mean free path is defined as $\ell^{*}=1/\mu_{s}'$.
	
	\subsection{Extrapolated (partial‑current) boundary conditions}
	\label{sec:EBC}
	At the physical interfaces $z=0$ and $z=L$ the net photon current must vanish
	when the exterior medium is non‑absorbing.  Within the diffusion framework this
	condition is expressed as a partial‑current (extrapolated) boundary condition
	\cite{Paasschens2008,CaseZweifel1967}:
	\begin{equation}
		\Phi(z,\lambda)+z_{b}(\lambda)\,\frac{\partial\Phi}{\partial n}=0
		\quad\text{at}\quad z=0,\;L,
		\label{eq:EBC}
	\end{equation}
	where the extrapolation length is
	\begin{equation}
		z_{b}=2A\,D,\qquad
		A=\frac{1+R_{\mathrm{eff}}}{1-R_{\mathrm{eff}}}.
	\end{equation}
	$R_{\mathrm{eff}}$ is the hemispherical internal reflectance evaluated for the
	effective refractive index of the composite (see Eq.~\ref{eq:neff} in Sec.~\ref{sec:material}).
	The condition (\ref{eq:EBC}) is mathematically equivalent to imposing zero
	fluence at virtual planes located at $z=-z_{b}$ and $z=L+z_{b}$.
	
	\subsection{Mapping external illumination to an internal isotropic source}
	\label{sec:sourceMapping}
	For a collimated beam of spectral photon irradiance $E_{\mathrm{in}}(\lambda)$
	incident on the front face, a fraction $R_{\mathrm{ext}}(\lambda)$ is reflected.
	The net injected photon flux is therefore
	\begin{equation}
		S_{0}(\lambda)=\big[1-R_{\mathrm{ext}}(\lambda)\big]\,E_{\mathrm{in}}(\lambda).
		\label{eq:S0}
	\end{equation}
	The first scattering event rapidly isotropises the photon direction.  In the
	diffusion regime this process can be represented by a thin isotropic plane source
	located at the average first‑scatter depth
	\begin{equation}
		z_{0}(\lambda)\approx\ell^{*}(\lambda)=\frac{1}{\mu_{s}'(\lambda)} .
		\label{eq:z0}
	\end{equation}
	Mathematically the source term becomes a Dirac delta distribution,
	\begin{equation}
		S(\mathbf{r},\lambda)=S_{0}(\lambda)\,\delta\!\big(z-z_{0}(\lambda)\big),
		\label{eq:source}
	\end{equation}
	which is the $z_{1}\!\to\!0$ limit of a uniform thin source layer of thickness
	$z_{1}$.  This representation is exact for a diffusion description because
	beyond $z_{0}$ the angular distribution is already isotropic.  Limitations
	arise only when a substantial ballistic component survives for $z\gtrsim\ell^{*}$,
	a situation that does not occur for the highly scattering aerogels studied
	here.
	
	\subsection{Finite‑slab Green’s function}
	\label{sec:greens}
	Substituting the source (\ref{eq:source}) into the diffusion equation
	(\ref{eq:diffusion}) and enforcing the extrapolated boundaries (\ref{eq:EBC})
	yields the Green’s function solution
	\begin{equation}
		\Phi(z,\lambda)=\frac{S_{0}(\lambda)}{D(\lambda)\,\kappa(\lambda)}\,
		\frac{\sinh\!\big[\kappa(\min\{z,z_{0}\}+z_{b})\big]\;
			\sinh\!\big[\kappa(L+z_{b}-\max\{z,z_{0}\})\big]}
		{\sinh\!\big[\kappa L_{e}\big]},
		\label{eq:PhiSlab}
	\end{equation}
	where $\kappa(\lambda)=\sqrt{\mu_{a}(\lambda)/D(\lambda)}$ and
	$L_{e}=L+2z_{b}$.  The derivation, including the homogeneous solution,
	matching at $z_{0}$, the jump condition for $\partial_{z}\Phi$, and the
	implementation of the extrapolated boundaries, is presented in full detail in
	Appendix~A.
	
	Integrating (\ref{eq:PhiSlab}) over the slab thickness gives the total fluence
	\begin{equation}
		\int_{0}^{L}\!\Phi(z,\lambda)\,dz=
		\frac{S_{0}(\lambda)}{\mu_{a}(\lambda)}\,
		\frac{\sinh\!\big[\kappa(z_{0}+z_{b})\big]\;
			\sinh\!\big[\kappa(L+z_{b}-z_{0})\big]}
		{\sinh\!\big[\kappa L_{e}\big]} .
		\label{eq:IntPhi}
	\end{equation}
	
	\subsection{Compact asymptotic predictor}
	
	Using Eqs.~\eqref{eq:PhiSlab}--\eqref{eq:IntPhi}, we choose $z_0=\ell^*$ for collimated beams and define $\langle \Phi \rangle = L^{-1} \! \int_0^L \Phi \, dz$. For $\mu_a \ll \mu_s'$ and modest $\kappa L_e$, the expansion yields the simple expression:
	\begin{equation}
		\boxed{
			\langle \Phi \rangle \approx \frac{S_0}{\mu_a L}\,\frac{\ell^*}{\ell^* + \frac{L}{2} + z_b}.
		}
		\label{eq:compactFluence}
	\end{equation}
	Equation (\ref{eq:compactFluence}) makes explicit the transition from the $\ell^{*2}/L$ scaling (strongly diffusive) to the $\ell^{*}$ scaling (optically thin) and retains the boundary correction $z_{b}/\ell^{*}$. It importantly constitutes a very compact design law, as will be seen hereafter.
	
	\section{Optical‑to‑kinetic mapping}
	\label{sec:opt2kin}
	
	Having obtained the photon fluence field $\Phi(z,\lambda)$ from the diffusion
	solution (\ref{eq:PhiSlab}), we now translate the optical quantities into
	photocatalytic reaction rates.  In photon‑unit
	conventions, the resulting kinetic constants are expressed per incident photon.
	
	The local photon absorption rate density is
	\begin{equation}
		q_{a}(z,\lambda)=\mu_{a}(\lambda)\,\Phi(z,\lambda)\qquad
		\big[\si{\photon\per\cubic\meter\per\second}\big].
	\end{equation}
	Only a fraction $\phi_{\mathrm{int}}(\lambda)$ of the absorbed photons creates
	usable charge carriers (electron–hole pairs or radicals).  This internal quantum
	efficiency is defined as
	\begin{equation}
		\phi_{\mathrm{int}}(\lambda)=\frac{\text{useful carriers generated}}
		{\text{photons absorbed}},
	\end{equation}
	and typically lies between 0.1 and 0.3 for anatase TiO$_2$ under UV
	illumination \cite{deBoer2019}.  The primary carrier generation
	rate density therefore reads
	\begin{equation}
		G(z,\lambda)=\phi_{\mathrm{int}}(\lambda)\,\mu_{a}(\lambda)\,\Phi(z,\lambda).
		\label{eq:Gdef}
	\end{equation}
	
	Photocatalytic reactions take place at the solid–fluid interface.  We denote
	by $S_{\mathrm{accessible}}$ the accessible catalytic surface per macroscopic
	volume (units \si{\per\meter}) and by $k_{\mathrm{surf}}$ the first‑order
	surface rate coefficient (units \si{\meter}).  Assuming a site‑limited
	mechanism, the local volumetric reaction rate is \cite{Hoffmann1995}
	\begin{equation}
		r_{V}(z,\lambda)=k_{\mathrm{surf}}\;S_{\mathrm{accessible}}\;G(z,\lambda) .
		\label{eq:rVsurf}
	\end{equation}
	
	Integrating Eq.~\ref{eq:rVsurf} over the slab thickness $L$ yields the total
	volumetric rate per illuminated area $A$,
	\begin{equation}
		\frac{R_{V}(\lambda)}{A}=k_{\mathrm{surf}}\;S_{\mathrm{accessible}}\;
		\phi_{\mathrm{int}}(\lambda)\,\mu_{a}(\lambda)
		\int_{0}^{L}\!\Phi(z,\lambda)\,\mathrm{d}z .
		\label{eq:RV}
	\end{equation}

	The intrinsic volumetric rate constant per incident photon is defined as
	\begin{equation}
		k_{V,\mathrm{mono}}(\lambda)\equiv
		\frac{R_{V}(\lambda)/A}{S_{0}(\lambda)}
		= k_{\mathrm{surf}}\;S_{\mathrm{accessible}}\;
		\phi_{\mathrm{int}}(\lambda)\,
		\frac{\displaystyle\int_{0}^{L}\!\Phi(z,\lambda)\,\mathrm{d}z}
		{\displaystyle S_{0}(\lambda)} .
		\label{eq:kVmono_def}
	\end{equation}
	
	Substituting the Green’s‑function solution for the fluence
	(Eq.~\ref{eq:PhiSlab}), is integrated form (Eq.~\ref{eq:IntPhi}) and the source term $S_{0}(\lambda)$ (Eq.~\ref{eq:S0})
	gives the exact diffusion‑approximation expression for a single wavelength:
	\begin{equation}
		k_{V,\mathrm{mono}}(\lambda)=
		k_{\mathrm{surf}}\;S_{\mathrm{accessible}}\;
		\phi_{\mathrm{int}}(\lambda)\,
		\frac{\sinh\!\bigl[\kappa(z_{0}+z_{b})\bigr]\;
			\sinh\!\bigl[\kappa(L+z_{b}-z_{0})\bigr]}
		{\sinh\!\bigl[\kappa L_{e}\bigr]} .
		\label{eq:kVexact}
	\end{equation}
	
	When the surface reaction follows a Langmuir–Hinshelwood isotherm, the
	kinetic prefactor acquires the coverage factor
	$\theta=K C/(1+K C)$ (with equilibrium constant $K$ and bulk concentration
	$C$).  In that case $k_{\mathrm{surf}}$ in Eq.~\ref{eq:kVexact} is simply
	replaced by $k_{\mathrm{LH}}\theta$, yielding an analogous expression for the
	intrinsic volumetric constant.
	
	In the diffusion–dominated regime ($\ell^{*}\!\ll\!L$) and for modest boundary
	corrections ($z_{b}\!\ll\!\ell^{*}$), the depth-integrated fluence
	$\int_{0}^{L}\Phi(z,\lambda)\,\mathrm{d}z$ admits a compact, fully
	non–dimensionalised form (\ref{eq:compactFluence}).  Using
	$S_{\mathrm{accessible}}=\chi_{\mathrm{access}}(S/V)$, where $S/V$ is the
	macroscopic surface–to–volume ratio and $\chi_{\mathrm{access}}$ measures the
	fraction of that surface that is catalytically accessible, the intrinsic
	volumetric rate constant per incident photon becomes
	\begin{equation}
		\boxed{
			k_{V,\mathrm{mono}}(\lambda)
			= k_{\mathrm{surf}}\,\chi_{\mathrm{access}}\,
			\phi_{\mathrm{int}}(\lambda)\,
			\frac{(S/V)\,\ell^{*}}
			{1+\dfrac{L}{2\ell^{*}}+\dfrac{z_{b}}{\ell^{*}}} }.
	\label{eq:kVcompact}
	\end{equation}
	This expression corresponds to a \emph{site-limited surface-reaction picture}:
	photons generate charge carriers in the bulk, and these carriers react at
	accessible surface sites with an effective first-order coefficient
	$k_{\mathrm{surf}}$.  
	
	An alternative but mathematically parallel viewpoint is obtained by describing
	the surface in terms of discrete active sites of areal density $N_{\mathrm{site}}$,
	each characterised by an absorption cross section $\sigma_{\mathrm{abs}}(\lambda)$.
	In this \emph{photon-absorption-limited picture}, the fraction of incident
	photons captured by catalytic sites is $\sigma_{\mathrm{abs}}N_{\mathrm{site}}$,
	and replacing $k_{\mathrm{surf}}\chi_{\mathrm{access}}$ by this optical
	prefactor yields the intrinsic areal rate constant
	\begin{equation}
		\boxed{
			k_{\mathrm{photo},A}(\lambda)
			= \phi_{\mathrm{int}}(\lambda)\,
			\sigma_{\mathrm{abs}}(\lambda)\,
			N_{\mathrm{site}}\,
			\frac{(S/V)\,\ell^{*}}
			{1+\dfrac{L}{2\ell^{*}}+\dfrac{z_{b}}{\ell^{*}}} }.
	\label{eq:kAcompact}
	\end{equation}
	Both compact forms share the same transport factor—capturing diffusion,
	geometry, and boundary effects—while differing only in their microscopic
	reaction prefactor.  The volumetric expression is appropriate when surface
	kinetics limit the reaction, whereas the areal expression is appropriate when
	photon absorption by surface sites is rate-determining.

	In practice the workflow proceeds as follows: the measured (or retrieved)
	optical coefficients $\mu'_{s}(\lambda)$, $\mu_{a}(\lambda)$ and anisotropy
	factor $g(\lambda)$ are used to compute the diffusion parameters $D$,
	$\kappa$, $z_{b}$ and the source depth $z_{0}\approx\ell^{*}$.  The Green’s‑function
	solution (Eq.~\ref{eq:PhiSlab}) provides the fluence $\Phi(z,\lambda)$, which
	is converted to the carrier‑generation rate $G(z,\lambda)$ via Eq.~\ref{eq:Gdef}.
	Insertion of $G$ into the kinetic expression (Eq.~\ref{eq:rVsurf}) and depth
	integration yields the exact intrinsic constant $k_{V,\mathrm{mono}}(\lambda)$
	(Eq.~\ref{eq:kVexact}). The compact predictors (Eq.~\ref{eq:kVcompact} and Eq.~\ref{eq:kAcompact})
	can finally be employed for rapid design studies.
	
	\section{Core–Shell aerogel geometry and material‑parameter extraction}
	\label{sec:material}
	
\begin{figure}[t]
		\centering
		
		\begin{subfigure}[b]{0.45\textwidth}
			\centering
			\includegraphics[width=\linewidth]{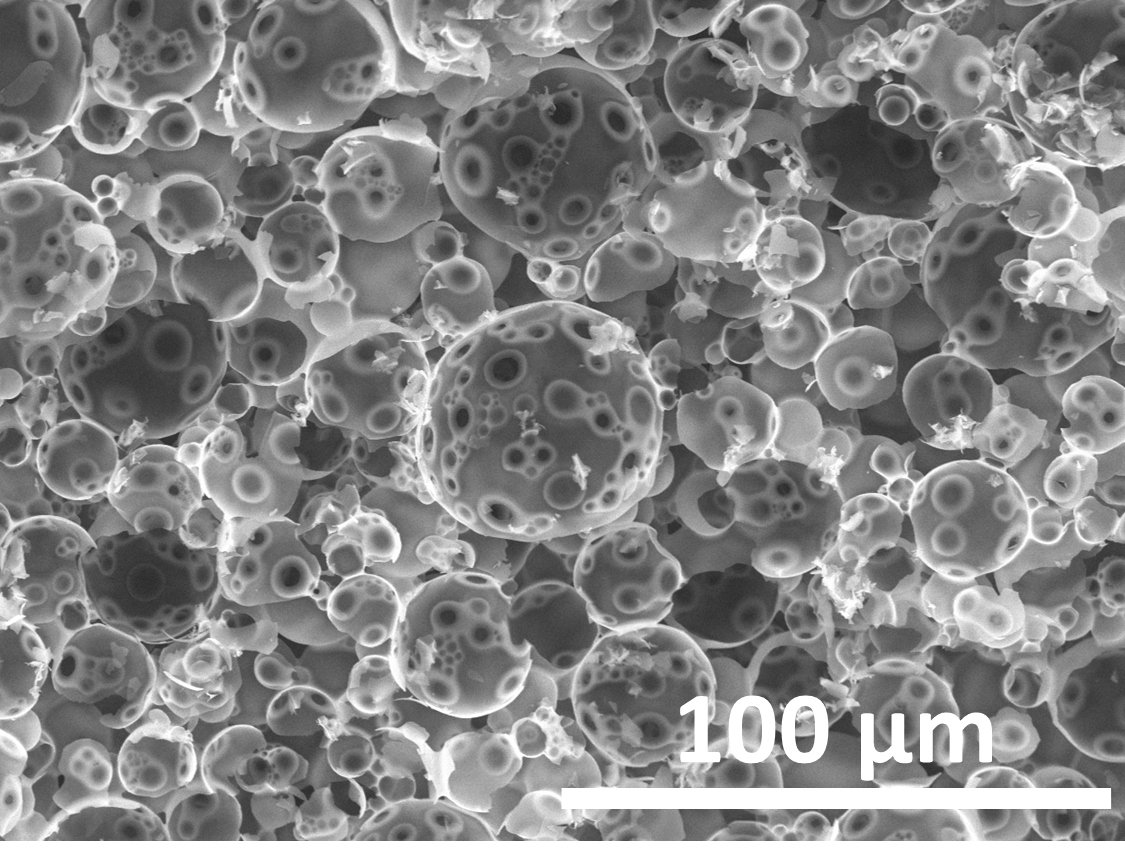}
			\caption{Macro-pores observed in SEM investigations at the macroscopic scale. Representative Si(HIPE)1000 sample of Ref.~\cite{Vardon2025}.}
			\label{fig:macropores}
		\end{subfigure}
		\hfill
		\begin{subfigure}[b]{0.45\textwidth}
			\centering
			\begin{tikzpicture}
				\node[inner sep=0pt] (main) {
					\includegraphics[width=\linewidth]{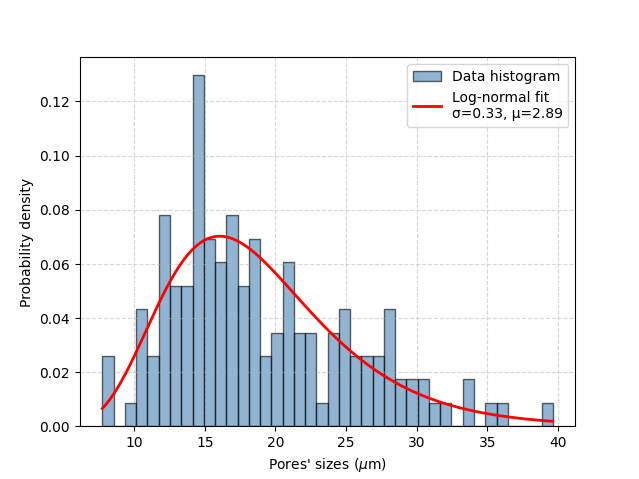}
				};
				\node[anchor=east, inner sep=0pt,
				xshift=-0.03\linewidth, yshift=0.03\linewidth] at (main.east) {
					\includegraphics[width=0.30\linewidth]{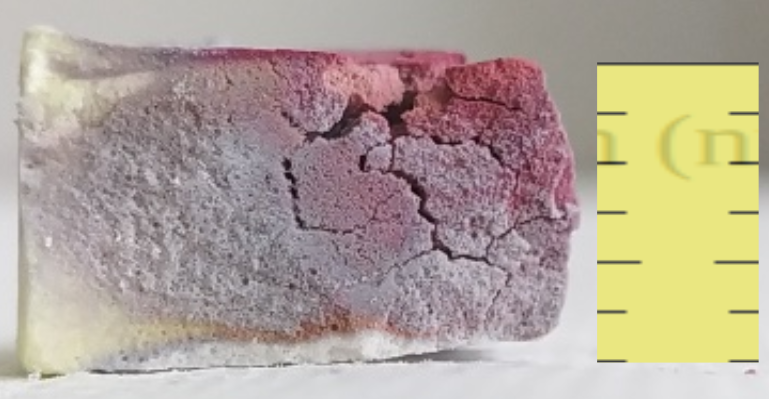}
				};
			\end{tikzpicture}
			\caption{Probability distribution function of the macro-pore sizes extracted from the representative Si(HIPE)1000 sample of Ref.~\cite{Vardon2025}, with a schematic view of light penetration depth in the monolith.}
			\label{fig:pdf}
		\end{subfigure}
		
		\caption{Macroporous morphology and statistical pore-size distribution of Si(HIPE)1000 samples.}
	\end{figure}

	The optical coefficients that enter the diffusion model—reduced scattering
	$\mu'_{s}(\lambda)$, absorption $\mu_{a}(\lambda)$, and anisotropy factor $g(\lambda)$—are obtained from the hierarchical
	core–shell aerogel macrostructure (Figure~\ref{fig:macropores}). Highly idealized, the geometry consists of air‑filled spherical
	pores (core) of radius $R_{1}\approx\SI{18}{\micro\meter}$ (Figure~\ref{fig:pdf}) embedded in a silica
	matrix (shell) that contains anatase TiO$_{2}$ nanoparticles (NPs) of radius
	$a_{\mathrm{np}}\approx\SI{7.45}{\nano\meter}$.  The shell thickness is
	$t\approx\SI{70}{\nano\meter}$ and the overall porosity $\phi\approx =0.9$
	\cite{Vardon2025}.
	
	\subsection{Effective refractive index of the shell}
	For a dilute inclusion concentration the Maxwell–Garnett mixing rule
	\cite{MaxwellGarnett1904} is generally employed:
	\begin{equation}
		\varepsilon_{\mathrm{shell}}=
		\varepsilon_{\mathrm{silica}}\!
		\left[
		1+3\,\frac{f_{\mathrm{NP}}\,
			\bigl(\varepsilon_{\mathrm{NP}}-\varepsilon_{\mathrm{silica}}\bigr)}
		{\varepsilon_{\mathrm{NP}}+2\varepsilon_{\mathrm{silica}}
			-f_{\mathrm{NP}}\bigl(\varepsilon_{\mathrm{NP}}-\varepsilon_{\mathrm{silica}}\bigr)}
		\right],
		\label{eq:MG}
	\end{equation}
	where $f_{\mathrm{NP}}$ is the TiO$_{2}$ nanoparticle volume fraction,
	$\varepsilon_{\mathrm{NP}}$ and $\varepsilon_{\mathrm{silica}}$ are the
	dielectric functions of TiO$_{2}$ and silica, respectively.
	
	When $f_{\mathrm{NP}}$ exceeds $\sim0.05$, the interaction between inclusions
	becomes non‑negligible and the Bruggeman symmetric EMA
	\cite{Bruggeman1935} might be more appropriate:
	\begin{equation}
		f_{\mathrm{NP}}\frac{\varepsilon_{\mathrm{NP}}-\varepsilon_{\mathrm{shell}}}
		{\varepsilon_{\mathrm{NP}}+2\varepsilon_{\mathrm{shell}}}
		+(1-f_{\mathrm{NP}})\frac{\varepsilon_{\mathrm{silica}}-
			\varepsilon_{\mathrm{shell}}}{\varepsilon_{\mathrm{silica}}+2\varepsilon_{\mathrm{shell}}}=0 .
		\label{eq:Bruggeman}
	\end{equation}
	For our geometry, both models give virtually identical $n_{\mathrm{shell}}$ (differences $<2\%$)
	over the $f_{\mathrm{NP}}$ range explored (see Supplementary json file collecting parameters for various wavelengths). Accordingly, we use exclusively the Maxwell–Garnett mixing rule in the whole study.
	
	The effective refractive index entering the boundary condition is then
	\begin{equation}
		n_{\mathrm{eff}}=\sqrt{\phi\,\varepsilon_{\mathrm{air}}+(1-\phi)\,
			\varepsilon_{\mathrm{shell}}}\; .
		\label{eq:neff}
	\end{equation}
	
	\subsection{Core–shell Mie scattering}
	The macroscopic scattering properties are obtained from full core–shell Mie
	calculations using the Aden–Kerker formalism \cite{AdenKerker1951}.  The
	outer radius of a core–shell particle is $a=R_{1}+t$, and the effective shell
	index $n_{\mathrm{shell}}$ has been taken in this paper (except for the sake of comparison) from Eq.~\ref{eq:MG}.
	For each wavelength the Mie code returns
	the scattering efficiency $Q_{s}(\lambda)$, absorption efficiency $Q_{a}(\lambda)$,
	and the asymmetry factor $g(\lambda)$.
	
	All Mie calculations were performed with the open‑source Python package
	\texttt{PyMieScatt} \cite{Sumlin2018,PyMieScatt}, which implements the Aden–Kerker
	expressions and provides rigorous checks of energy conservation.
	
	\subsection{Reduced scattering coefficient}
	The number density of the macropores follows from the porosity $\phi$ and the
	core radius $R_{1}$:
	\begin{equation}
		N_{\mathrm{core}}=\frac{\phi}{\tfrac{4}{3}\pi R_{1}^{3}} .
		\label{eq:Ncore}
	\end{equation}
	The geometric scattering cross‑section of a single core–shell particle is
	$\sigma_{s}=\pi a^{2}Q_{s}(\lambda)$, giving
	\begin{equation}
		\mu_{s}(\lambda)=N_{\mathrm{core}}\;\sigma_{s}(\lambda),\qquad
		\mu_{s}'(\lambda)=\mu_{s}(\lambda)\,[1-g(\lambda)] .
		\label{eq:mus}
	\end{equation}
	The transport mean free path is $\ell^{*}=1/\mu_{s}'$.
	
	\subsection{Absorption coefficient from TiO$_{2}$ nanoparticles}
	The volume of TiO$_{2}$ contained in the shell of a single macroparticle is
	\begin{equation}
		V_{\mathrm{NP|shell}}=f_{\mathrm{NP}}\,
		\frac{4}{3}\pi\big[(R_{1}+t)^{3}-R_{1}^{3}\big] .
	\end{equation}
	Dividing by the nanoparticle volume $V_{\mathrm{NP}}=\tfrac{4}{3}\pi a_{\mathrm{np}}^{3}$
	gives the number of NPs per shell,
	$N_{\mathrm{NP|shell}}=V_{\mathrm{NP|shell}}/V_{\mathrm{NP}}$, and the bulk NP
	number density
	\begin{equation}
		n_{\mathrm{NP}}=N_{\mathrm{core}}\;N_{\mathrm{NP|shell}} .
	\end{equation}
	The absorption coefficient is then
	\begin{equation}
		\mu_{a}(\lambda)=f_{\mathrm{abs,eff}}\;n_{\mathrm{NP}}\;
		\sigma_{\mathrm{abs}}(\lambda) ,
		\label{eq:mua}
	\end{equation}
	where $\sigma_{\mathrm{abs}}=\pi a_{\mathrm{np}}^{2}Q_{a}$ and
	$f_{\mathrm{abs,eff}}$ accounts for local‑field and accessibility effects.

	\subsection{Phase‑function for anisotropic scattering}
	When an explicit phase function is required (e.g. for Monte‑Carlo benchmarks)
	the Henyey–Greenstein form \cite{Henyey1941} is adopted:
	\begin{equation}
		p(\cos\theta)=\frac{1-g^{2}}{\bigl(1+g^{2}-2g\cos\theta\bigr)^{3/2}} .
		\label{eq:HG}
	\end{equation}
	The anisotropy factor $g(\lambda)$ is taken directly from the Mie
	calculations (values $0.848$–$0.864$ for the investigated packing fractions).
	
	\subsection{Summary of material parameters}
	Table~\ref{tab:material} gathers the key quantities that feed the diffusion‑kinetic
	framework for the representative packing fractions.  All values are derived
	from the procedures outlined above and are used in the subsequent rate‑constant
	calculations.
	
	\begin{table}[H]
		\centering
		\caption{Representative material parameters (reference wavelength
			$\lambda=\SI{0.340}{\micro\meter}$).  $n_{\mathrm{shell}}$ is obtained
			from Eq.~\ref{eq:MG}.  $g$, $\mu_{s}'$, $\mu_{a}$,
			and $\ell^{*}=1/\mu_{s}'$ follow from the core–shell Mie calculations.}
		\begin{tabular}{cccccc}
			\toprule
			$f_{\mathrm{NP}}$ & $n_{\mathrm{shell}}$ & $g$ &
			$\mu_{s}'$ (\si{\per\meter}) &
			$\mu_{a}$ (\si{\per\meter}) &
			$\ell^{*}$ (\si{\micro\meter}) \\
			\midrule
			0.01 & 1.45 & 0.848 & $1.08\times10^{4}$ & $3.63\times10^{2}$ & 92 \\
			0.05 & 1.50 & 0.850 & $1.17\times10^{4}$ & $1.81\times10^{3}$ & 85 \\
			0.10 & 1.57 & 0.854 & $1.26\times10^{4}$ & $3.63\times10^{3}$ & 80 \\
			0.20 & 1.71 & 0.864 & $1.36\times10^{4}$ & $7.26\times10^{3}$ & 74 \\
			\bottomrule
		\end{tabular}
		\label{tab:material}
	\end{table}
	
	The extracted optical coefficients are inserted directly into the Green’s‑function
	solution (\ref{eq:PhiSlab}) and, via the mapping described in
	Section~\ref{sec:opt2kin}, yield the intrinsic volumetric and areal rate
	constants for any desired illumination spectrum.
	
	\section{Validation and uncertainty}
	\label{sec:validation}
	
	In this section we (i) assess the applicability of the diffusion approximation
	for the highly‑scattering aerogels, (ii) validate the finite‑slab diffusion solution against a
	rigorous Monte‑Carlo photon‑migration benchmark,  and (iii)
	propagate uncertainties in the optical and kinetic parameters to the final
	intrinsic rate constants.
	
	\subsection{Diffusion‑approximation criteria}
	Two dimensionless criteria are commonly used to justify the diffusion
	approximation \cite{Paasschens2008}:
	
	\begin{enumerate}[label=(\alph*)]
		\item \textbf{Transport‑mean‑free‑path criterion:}
		$\ell^{*}\ll L$.  For all packing fractions studied
		$\ell^{*}/L\in[0.0074,0.0092]$, comfortably satisfying this condition.
		
		\item \textbf{Anisotropy criterion:}
		$g\,\ell^{*}/L\lesssim0.01$.  Our Mie calculations give $g\in[0.848,0.864]$,
		leading to $g\,\ell^{*}/L\in[6.2\times10^{-3},7.9\times10^{-3}]$, well within the
		diffusion regime.
	\end{enumerate}
	
	\subsection{Monte‑Carlo photon‑migration benchmark}
	To quantify the accuracy of the diffusion solution we performed one‑dimensional
	Monte‑Carlo simulations (algorithm reproduced in full in~\ref{app:MC}).  For 
	each nanoparticle packing fraction $f_{\mathrm{NP}}$
	we launched $10^{8}$ photon packets, sampled free‑path lengths from the
	exponential distribution with total extinction coefficient
	$\mu_{t}=\mu_{a}+\mu_{s}$, and scattered them according to the
	Henyey–Greenstein phase function (Eq.~\ref{eq:HG}) using the anisotropy factor
	$g$ obtained from the core–shell Mie calculations.  Boundary interactions
	employed the internal hemispherical reflectance $R_{\mathrm{eff}}$ computed from
	the effective refractive index $n_{\mathrm{eff}}$ (Eq.~\ref{eq:neff}).
	
\begin{figure}[H]
		\centering
		\includegraphics[width=0.85\linewidth]{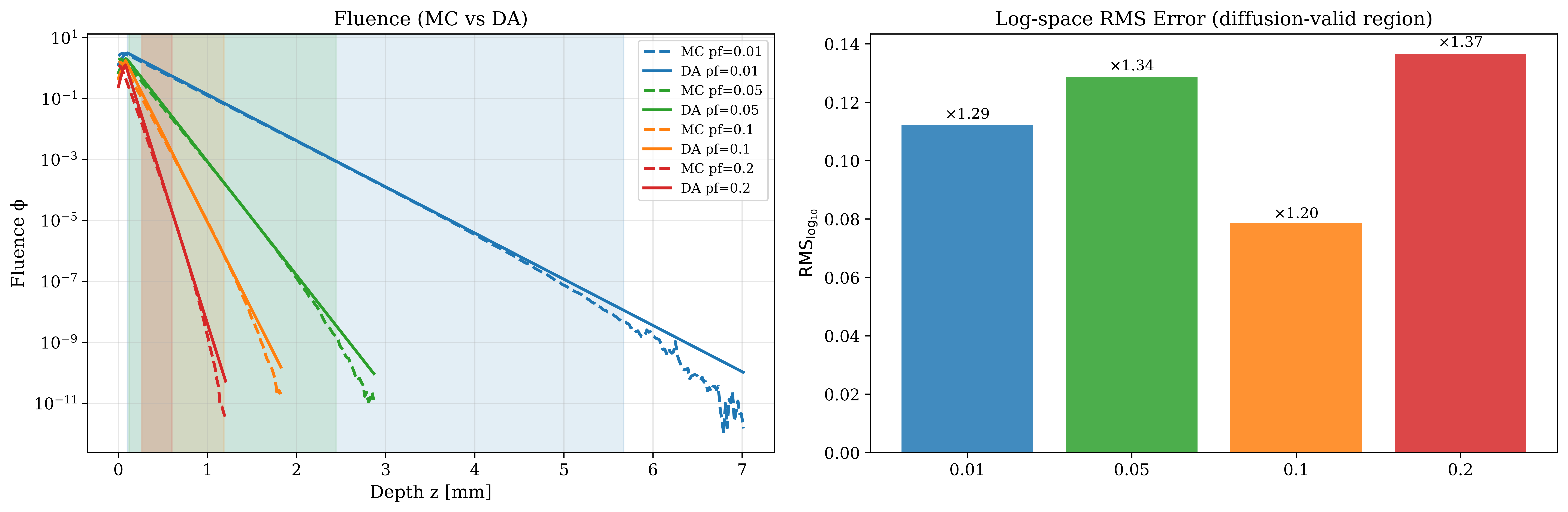}
		\caption{Left: Depth‑dependent fluence $\Phi(z)$ for the four representative packing fractions.
			Solid line: analytical diffusion solution (Eq.~\ref{eq:compactFluence});
			Dash line: Monte‑Carlo photon‑migration simulation (~\ref{app:MC}).
			The right panel shows the relative $\mathrm{RMS}_{\log_{10}}$.}
		\label{fig:MCvsAnalytical}
	\end{figure}

	Figure~\ref{fig:MCvsAnalytical} compares the Monte‑Carlo fluence
	$\Phi_{\text{MC}}(z)$ with the analytical Green’s‑function solution
	(Eq.~\ref{eq:compactFluence}) for the four representative packing fractions.
	
	To quantitatively compare the Monte Carlo (MC) fluence with the diffusion
	approximation (DA), we first removed the non-physical zero tail of the MC
	data and performed all analyses on the physically meaningful range only.
	The diffusion-valid region was then identified automatically using the
	curvature of the log-scale MC fluence. In diffusion theory the fluence
	decays exponentially, and therefore appears as a straight line when plotted
	as $\log_{10}\phi(z)$ versus depth. Deviations from linearity indicate
	either pre-diffusive transport near the source or noise-dominated behaviour
	in the deep tail. We computed the second derivative of $\log_{10}\phi(z)$,
	smoothed it, and extracted the longest contiguous interval where the
	curvature remained below a fixed relative threshold. This interval defines
	the region in which the DA can be meaningfully compared to MC.
	
	Within this automatically detected diffusion-valid region, we evaluated the
	difference between DA and MC in logarithmic space. The corresponding
	root-mean-square error,
	\[
	\mathrm{RMS}_{\log_{10}}
	= \sqrt{\left\langle
		\left( \log_{10}\phi_{\mathrm{DA}}
		- \log_{10}\phi_{\mathrm{MC}}
		\right)^2 \right\rangle},
	\]
	provides a scale-invariant and physically interpretable metric: a value of
	$\mathrm{RMS}_{\log_{10}} = \Delta$ corresponds to a typical multiplicative
	deviation of $10^{\Delta}$ between DA and MC.
	
	The results show that the diffusion approximation reproduces the MC fluence
	reasonably well across all investigated phase functions. In the
	diffusion-valid region, the multiplicative DA--MC mismatch remains modest,
	with typical factors ranging from approximately $1.20$ to $1.39$ depending
	on the scattering phase function. As expected, the agreement degrades
	slightly for more strongly forward-peaked scattering (larger $p_f$), yet
	the DA still captures the overall depth dependence of the fluence with
	surprisingly good accuracy. These values confirm that, once restricted to
	its proper regime of validity, the diffusion approximation provides a
	robust and quantitatively reliable representation of the fluence predicted
	by full Monte Carlo simulation.
	
	\subsection{Uncertainty propagation}
	
	At the highest TiO$_2$ loading ($f_{\mathrm{NP}} = 0.20$), the dependent-scattering correction of Lax~\cite{Lax1952,Vynck2023} corresponds to a global
	nanoparticle filling fraction $\beta \approx 2\times10^{-3}$, i.e.\ a reduction of
	only $\approx 0.2\%$ in $\mu_s'$. Propagating this correction through the full
	diffusion expression (Eq.~\ref{eq:kVexact}) modifies the intrinsic volumetric rate
	constant $k_{V,\mathrm{mono}}$ by less than $10^{-3}$. A detailed derivation, together
	with a discussion of the conditions under which the Lax correction becomes non-negligible, is provided in~\ref{app:Lax}.
	
	Propagating independent 5\% relative uncertainties in the relevant input
	parameters through Eq.~\ref{eq:kVcompact} and ~\ref{eq:kAcompact}, we find that uncertainties in the
	purely multiplicative prefactors ($k_{\mathrm{surf}}$, $\chi_{\mathrm{access}}$,
	$\phi_{\mathrm{int}}$ and $S_{\mathrm{accessible}}$ for $k_{V,\mathrm{mono}}$;
	$\phi_{\mathrm{int}}$, $\sigma_{\mathrm{abs}}$, $N_{\mathrm{sites}}$ and
	$S_{\mathrm{accessible}}$ for $k_{\mathrm{photo},A}$) dominate the error budget,
	while the geometric diffusion parameters ($L$ and $z_b$) have a much weaker
	incidence. Overall, the resulting propagated relative uncertainties are of
	order $10\%$ (see ~\ref{app:sens} for details).
	
	As a consequence of this section, the compact predictors (Eq.~\ref{eq:kVcompact} and Eq.~\ref{eq:kAcompact}) can be used
	confidently as a fast, physics‑based design tool for optimizing photocatalytic
	slab geometries and compositions.
	
	\section{Results and discussion}
	\label{sec:results}
	
	\subsection{The whole picture}
	Applying the analytical framework described in Sections~\ref{sec:theory}–%
	\ref{sec:opt2kin} to the core–shell aerogel geometry of Sec.~\ref{sec:material},
	we obtain the intrinsic volumetric and areal photocatalytic rate constants as
	functions of the nanoparticle packing fraction $f_{\mathrm{NP}}$.  
	All calculations were performed at the reference
	wavelength $\lambda=\SI{0.340}{\micro\meter}$ (the peak absorption of anatase
	TiO$_2$).
	
\begin{figure}[H]
		\centering
		\includegraphics[width=0.85\linewidth]{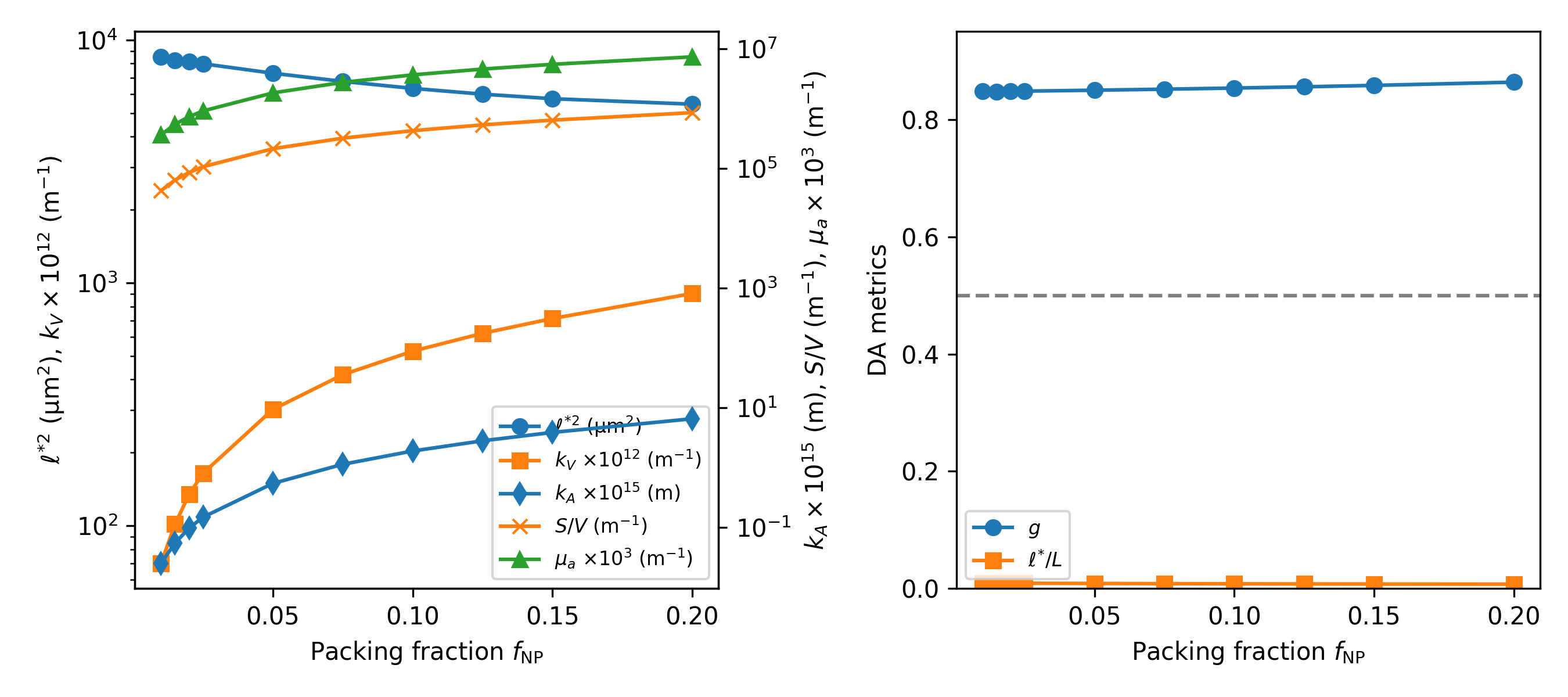}
		\caption{Evolution of optical, geometric and diffusion-approximation (DA) metrics 
			as a function of the TiO$_2$ nanoparticle packing fraction $f_{\mathrm{NP}}$.
			\textbf{Left:} Transport mean free path squared $\ell^{*2}$, intrinsic volumetric
			rate constant $k_V$, surface-based rate constant $k_A$, accessible 
			surface-to-volume ratio $S/V$, and absorption coefficient $\mu_a$. 
			While $\ell^{*2}$ shows only weak variations, all photonic activity metrics 
			($k_V$, $k_A$, $\mu_a$, and $S/V$) increase monotonically with nanoparticle
			loading owing to the thicker TiO$_2$ shell and enhanced absorption.
			\textbf{Right:} Diffusion-approximation validity metrics. The correction 
			factor $g$ remains close to $0.86$, and the ratio $\ell^*/L$ stays orders 
			of magnitude below the DA breakdown limit (dashed line), confirming that 
			the diffusion approximation remains valid throughout the full range of 
			$f_{\mathrm{NP}}$ studied}
		\label{fig:main}
	\end{figure}

	The trends observed in Fig.~\ref{fig:main} can be directly related
	to the effective optical parameters listed in Table~\ref{tab:material}.
	
	Figure~\ref{fig:main} shows how these effective parameters translate
	into global photonic and geometric metrics for the macroporous slab.  
	The absorption coefficient $\mu_a$ increases monotonically with
	$f_{\mathrm{NP}}$, and this trend is mirrored by both the intrinsic volumetric
	rate constant $k_V$ and the surface--based rate constant $k_A$, which quantify
	the volumetric and interfacial photogeneration efficiencies, respectively.  
	The accessible surface--to--volume ratio $S/V$ also increases with
	$f_{\mathrm{NP}}$, but for a different reason: as the silica shell is filled
	with a larger number of TiO$_2$ nanoparticles, each particle contributes its
	own nanoscale reactive surface area.  
	The total accessible surface therefore grows approximately in proportion to
	the number of inclusions, while the macropore volume changes only weakly,
	leading to a steady increase of $S/V$ with $f_{\mathrm{NP}}$.  
	In contrast, the transport parameter $\ell^{*2}$ varies only mildly, and the
	diffusion--approximation indicators $g$ and $\ell^{*}/L$ remain essentially
	constant, with $g\approx0.85$ and $\ell^{*}/L \ll 1$ across the entire
	composition range.  
	Taken together, these observations show that increasing $f_{\mathrm{NP}}$ up to
	$0.20$ strongly enhances absorption and reactive surface area (and thus $k_V$
	and $k_A$), while leaving the overall transport regime unchanged and the
	diffusion approximation safely valid.
	
	These trends can be interpreted mechanistically by recognising the two
	distinct levels of description in our model.  
	At the smallest scale, the TiO$_2$ nanoparticles are treated as inclusions
	dispersed within a silica host, and their cumulative effect on the local
	dielectric function is captured by an effective medium approximation.  
	Increasing $f_{\mathrm{NP}}$ at this level primarily increases the effective
	refractive index $n_{\mathrm{shell}}$ and the effective absorption coefficient
	inside the shell, as reported in Table~\ref{tab:material}.  
	At the next scale, each macropore wall is described as a large core--shell
	particle, with a low--index core and a shell of effective index
	$n_{\mathrm{shell}}$, and its scattering properties are obtained from Mie
	theory in the Aden--Kerker formulation.  
	Because the core--shell radius is much larger than the wavelength, these
	particles scatter light strongly in the forward direction, which explains the
	high and nearly constant anisotropy factors $g\approx0.85$.  
	The modest increase in $\mu_s'$ with $f_{\mathrm{NP}}$ then reflects the fact
	that, although the shell index contrast and number of inclusions increase, the
	macroscopic transport remains dominated by the large core--shell geometry and
	by forward--biased scattering.  
	In contrast, the substantial growth of $\mu_a$, $S/V$, $k_V$ and $k_A$
	directly follows from the increased density of absorbing TiO$_2$ inclusions
	and the larger total nanoparticle surface area available for interfacial
	photocatalysis.  
	Thus, Table~\ref{tab:material} and Fig.~\ref{fig:main} together
	illustrate how microscopic changes in the TiO$_2$ filling of the shell,
	captured by EMA, translate into macroscopic scattering and absorption
	properties at the core--shell level, and ultimately into enhanced photonic
	activity in the macroporous slab without compromising the validity of the
	diffusion approximation.
	
	\subsection{Two kinetic descriptions}
\begin{figure}[H]
		\centering
		\includegraphics[width=0.85\linewidth]{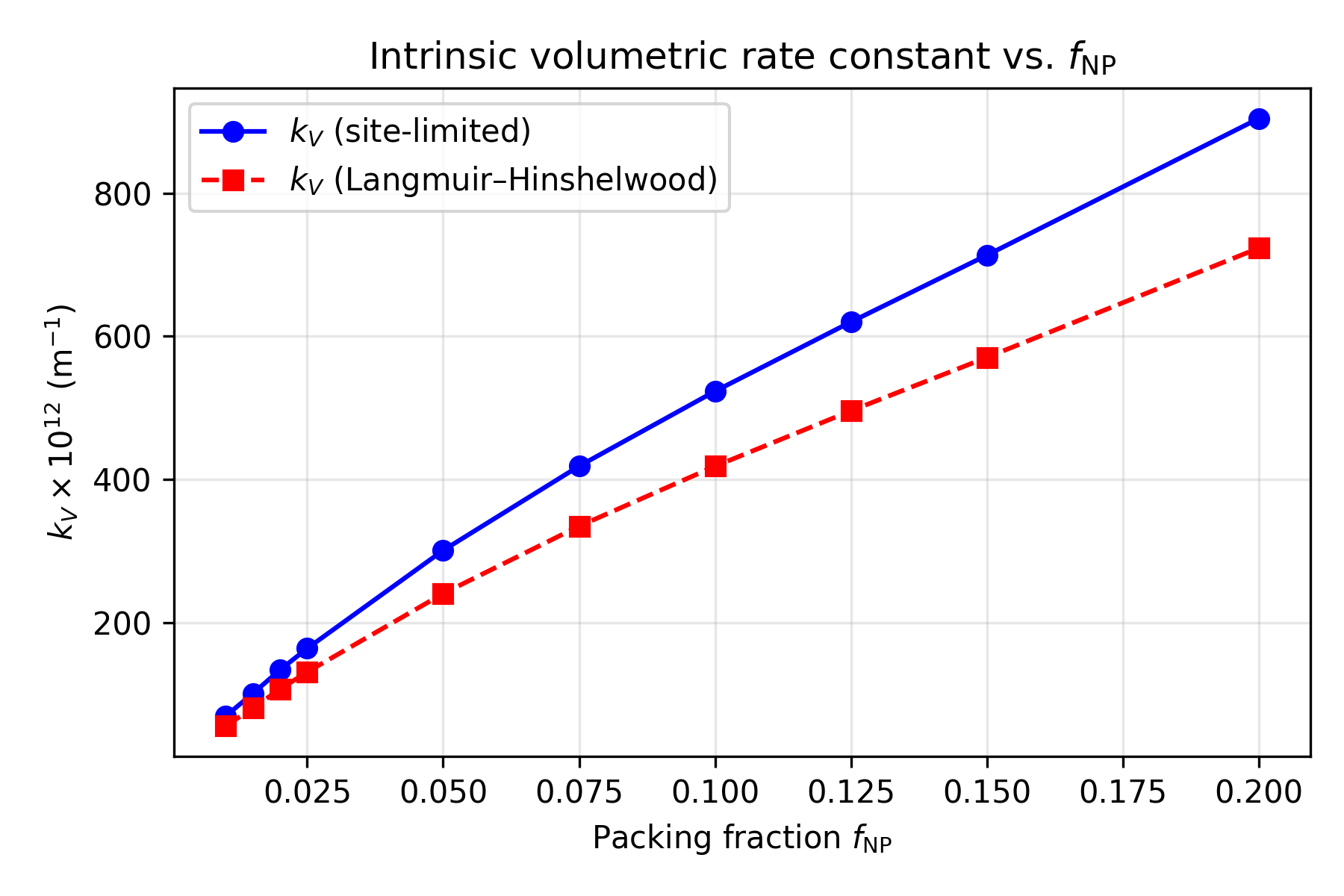}
		\caption{Intrinsic volumetric rate constant $k_{V,\mathrm{mono}}$ (solid
			lines) and compact predictor $k_{\mathrm{photo},A}$ (dashed lines) as a
			function of nanoparticle packing fraction $f_{\mathrm{NP}}$ for the
			site‑limited (blue) and Langmuir–Hinshelwood (red) kinetic models.
			Symbols denote the values obtained from the full Green’s‑function
			solution; the lines are guides to the eye.}
		\label{fig:kV}
	\end{figure}

	By inserting the fluence profiles into the kinetic mapping (\ref{eq:kVexact}), we here determine $k_{V,\mathrm{mono}}$
	as a function of $f_{\mathrm{NP}}$ (Fig.~\ref{fig:kV}) for the two kinetic
	descriptions here above considered:
	
	\begin{enumerate}[label=(\roman*)]
		\item \textbf{Site‑limited first‑order surface reaction} with
		$k_{\mathrm{surf}}=\SI{1e-9}{\meter}$.
		\item \textbf{Langmuir–Hinshelwood (LH) mechanism} with
		$k_{\mathrm{LH}}=\SI{5e-10}{\meter}$ and an equilibrium term
		$K C=0.6$.
	\end{enumerate}
	
	Figure~\ref{fig:kV} compares the intrinsic volumetric rate constants
	obtained for the two kinetic descriptions.  In both cases
	$k_{V,\mathrm{mono}}$ increases almost linearly with $f_{\mathrm{NP}}$, in line
	with the strong increase of the absorption coefficient and of the accessible
	surface-to-volume ratio $S/V$.  The Langmuir--Hinshelwood (LH) prediction
	(red dashed curve) lies systematically below the site-limited first-order
	result (blue solid curve), but follows essentially the same dependence on
	$f_{\mathrm{NP}}$.  For the chosen parameters ($k_{\mathrm{surf}}=10^{-9}\,\si{\meter}$,
	$k_{\mathrm{LH}}=5\times10^{-10}\,\si{\meter}$, $K C = 0.6$), the LH mechanism
	therefore behaves as an effectively first-order law with a reduced prefactor,
	reflecting the fact that only a fraction of the surface sites is free at any
	time.  A genuine LH saturation with respect to morphology would manifest as a
	sub-linear $k_{V,\mathrm{mono}}(f_{\mathrm{NP}})$ (the LH curve bending towards a
	plateau as additional TiO$_2$ surface is added), which is not observed here.
	Such saturation would require larger $K C$ or higher pollutant concentrations,
	so that the surface coverage approaches unity and extra TiO$_2$ surface area
	produces diminishing returns.
	
	\subsection{Design implications}
	
	The asymptotic expressions derived in Eqs.~(18) and (19) show that, in the
	strongly diffusive regime ($\ell^{*} \ll L$), both kinetic descriptions lead
	to a compact predictor of the form
	\begin{equation}
		k_{V,\mathrm{mono}} \;\simeq\;
		\mathcal{K}\,
		\frac{S/V}{L}\,\ell^{*2},
		\label{eq:kV_designlaw}
	\end{equation}
	where $\mathcal{K}$ collects the purely kinetic prefactors
	($k_{\mathrm{surf}}$ or $k_{\mathrm{LH}}$, $\phi_{\mathrm{int}}$, $\chi_{\mathrm{access}}$)
	and depends only weakly on morphology.  
	Equation~\eqref{eq:kV_designlaw} makes explicit that the photonic part of the
	problem is controlled by two dimensionless combinations,
	$S/V$ (how much reactive surface is available per unit volume) and
	$\ell^{*}/L$ (how deep photons can diffuse relative to the slab thickness).
	
	In the present aerogels the condition $\ell^{*} \ll L$ is very strongly
	satisfied: Table~\ref{tab:material} and Fig.~\ref{fig:main} show
	that $\ell^{*}/L$ remains in the narrow range $0.0092$--$0.0074$ for all
	$f_{\mathrm{NP}}$.  
	This confirms that light performs many scattering events before escaping,
	leading to a nearly homogeneous fluence field throughout the thickness.
	From a design perspective this is precisely what enables ``catalysis in the
	volume'': the reaction is not confined to a thin illuminated skin, but is
	driven by photons that have been redistributed over the entire macroporous
	network through multiple scattering.
	
	Within this deeply diffusive regime, Eq.~\eqref{eq:kV_designlaw} shows that
	increasing $k_{V,\mathrm{mono}}$ is essentially a matter of increasing (i) the
	available reactive surface $S/V$ and (ii) the characteristic transport length
	$\ell^{*}$, while keeping $\ell^{*}/L$ sufficiently small to maintain a
	volume-filling fluence.  
	The peculiar morphology of the present aerogels naturally pushes in this
	direction: the large fraction of voids and the core--shell architecture of the
	silica walls yield a high $S/V$, whereas the presence of scattering shells and
	structural imperfections keeps $\ell^{*}$ well below $L$ without driving it to
	extremely small values where transport would become too localised.  
	In other words, the macroporous network and its TiO$_2$-decorated shells act
	together as a ``multidiffuser'', spreading light efficiently through the
	volume while simultaneously providing a very large total nanoparticle surface.
	
	Equations~(18) and (19) therefore clarify the role of morphology.  
	They separate the purely geometric factor
	$(S/V)\,\ell^{*2}/L$ from the intrinsic kinetic constants, and show that
	optimising a highly scattering photocatalyst does not mean suppressing
	multiple scattering or making the slab arbitrarily thin.  
	Rather, the design target is an intermediate window where
	$\ell^{*}/L \ll 1$ (to ensure volumetric utilisation) but $\ell^{*}$ remains
	large enough that $(\ell^{*})^{2}/L$ and $S/V$ are both significant.  
	In this window---where the present aerogels operate---the large porosity and
	the physical nature of the scatterers (TiO$_2$-decorated shells, defects and
	imperfections) are not a drawback but the key enablers of efficient
	volume-based photocatalysis.

	\subsection{Extension to data‑driven design}
	\label{sec:dataDriven}
	
	Beyond providing compact design rules, the analytical framework developed in this work also opens a natural pathway toward data-driven optimisation of photocatalytic reactors. Recent studies have shown that machine-learning models can predict photocatalytic performance from a restricted set of physically meaningful descriptors. For example, Liu et al.~\cite{Liu2023} used gradient-boosted regression to estimate reaction rates from experimentally accessible features, demonstrating that accurate surrogate models can be trained even with limited datasets, provided the inputs encode the dominant optical and transport mechanisms.
	
	Data-driven design is likewise emerging in reactor engineering. Brunser et al.~\cite{Brunser2023} employed topology-optimised, additively manufactured ceramic lattices to tailor the spatial distribution of optical path lengths. By systematically varying the internal architecture, they effectively tuned transport-relevant quantities such as the photon mean free path~$\ell^{*}$ and reactor length~$L$, revealing performance trends consistent with transport-limited scaling arguments. Although their optimisation strategy did not incorporate the compact expressions derived here, their results highlight the practical value of analytical predictors—such as the $\ell^{*2}/L$ scaling law—for guiding geometry optimisation. Embedding such closed-form relations into topology-optimisation or machine-learning workflows would enable a far more targeted exploration of reactor architectures, substantially accelerating the discovery of high-performance photocatalytic designs.

	\subsection{Limitations and outlook}
	
	The present framework relies on two simplifying assumptions: a 
	wavelength-independent internal quantum efficiency $\phi_{\mathrm{int}}$ and 
	first-order surface kinetics. These assumptions allow us to derive compact, 
	closed-form expressions for the carrier generation rate and the volumetric 
	reaction rate, but they are not fundamental restrictions. In practice, 
	$\phi_{\mathrm{int}}$ can decrease toward longer wavelengths where carrier 
	recombination becomes more likely, and certain photocatalytic pathways may 
	exhibit higher-order or multi-electron kinetics. Extending the model to account 
	for a spectrally varying $\phi_{\mathrm{int}}(\lambda)$ or alternative kinetic 
	forms is conceptually straightforward: the analytical structure of the 
	derivation remains unchanged, with only the substitution of 
	$\phi_{\mathrm{int}}(\lambda)$ into Eq.~\ref{eq:Gdef} and a corresponding 
	generalisation of the surface‐reaction term in Eq.~\ref{eq:rVsurf}. Such 
	extensions would allow the compact predictors developed here to be applied to a 
	broader class of photocatalytic materials.
	
	The treatment of light transport also adopts the independent-scattering 
	approximation. Within the range of nanoparticle loadings considered, 
	comparisons with literature data indicate that dependent-scattering effects 
	modify the reduced scattering coefficient $\mu_{s}'$ by less than about 
	$8\%$. Although this lies well within the accuracy requirements of the present 
	study, future work could incorporate dependent-scattering corrections directly 
	into the optical model—for example via pair-distribution-function approaches or 
	structure-factor-modified Mie theory—to extend the applicability of the 
	framework to denser or strongly correlated nanocomposites.
	
	Overall, the analytical structure established in this work is readily adaptable. 
	Relaxing the assumptions on $\phi_{\mathrm{int}}$ and the reaction kinetics, 
	and enriching the optical model where necessary, will broaden the scope of the 
	framework while preserving its core advantage: physically transparent, 
	closed-form predictors that guide the design of high-performance photocatalytic 
	systems.

	\section{Conclusions}
	\label{sec:conclusions}
	
	This work establishes a compact analytical framework that links light transport,
	carrier generation, and surface reaction kinetics in highly scattering
	photocatalytic slabs. Using a finite-slab diffusion model with extrapolated
	boundaries, we obtain fluence fields that remain accurate even for strongly
	forward-scattering media ($g \approx 0.86$). Coupling these fields with a
	photochemical quantum efficiency and surface‐reaction kinetics yields closed-form
	expressions for the intrinsic volumetric and areal rate constants. These
	predictors capture, in a transparent manner, how performance depends on key
	transport and geometric parameters such as the transport mean free path, the
	effective optical thickness, and the surface-to-volume ratio.
	
	By condensing these dependencies into compact expressions
	(Eqs.~\ref{eq:kVcompact}–\ref{eq:kAcompact}), the framework provides a practical,
	low-cost tool for exploring the behaviour of diverse photocatalytic devices
	without resorting to full radiative-transfer simulations. The approach applies
	to systems where reaction takes place predominantly at surfaces as well as to
	configurations where a distributed, in-volume contribution may arise under
	appropriate optical conditions. This generality makes the framework relevant to
	numerous technologies studied in the literature—including solar-fuel reactors,
	VOC photo-degradation units, structured monoliths, aerogels, and
	additively-manufactured architectures.
	
	The formalism is also readily extensible. Although we assumed a constant
	internal quantum efficiency and first-order kinetics, these constraints can be
	relaxed by introducing a spectral $\phi_{\mathrm{int}}(\lambda)$ or alternative
	kinetic expressions without altering the structure of the derivation. Likewise,
	while the independent-scattering approximation is adequate for the loadings
	considered here, more advanced optical models could be incorporated to describe
	denser or structurally correlated nanocomposites.
	
	Beyond offering a predictive tool, the analytical expressions provide physically
	meaningful descriptors—such as $\ell^{*}$, $S/V$, and $z_b$—that interface
	naturally with data-driven design strategies and geometry-optimisation
	approaches. These descriptors help constrain large design spaces with
	physics-based structure, supporting experiment-efficient optimisation of
	photocatalytic devices across different operating principles and application
	domains.
	
	In summary, the compact predictors developed in this work supply a versatile and
	physically grounded foundation for analysing and designing photocatalytic slabs.
	Their combination of generality, transparency, and computational efficiency
	makes them applicable to a wide range of systems—from solar-fuel production to
	VOC photo-degradation and emerging additively-manufactured reactors—while also
	providing a platform for future data-driven and materials-optimisation
	strategies.

	\clearpage
	\appendix
	\renewcommand{\thesection}{Appendix~\Alph{section}}
	
	\section{Derivation of the finite‑slab Green’s function}
	\label{app:greens}
	
	We start from the steady‑state diffusion equation with a plane source
	(see Eq.~\ref{eq:diffusion}):
	
	\begin{equation}
	- D\,\frac{\mathrm{d}^{2}\Phi(z)}{\mathrm{d}z^{2}}+\mu_{a}\,\Phi(z)
	= S_{0}\,\delta\!\bigl(z-z_{0}\bigr) .
	\tag{A1}
	\end{equation}
	
	\subsection{Homogeneous solution}
	For $z\neq z_{0}$ the right‑hand side vanishes and we obtain the homogeneous
	ODE
	
	\begin{equation}
	\frac{\mathrm{d}^{2}\Phi}{\mathrm{d}z^{2}}-\kappa^{2}\Phi=0,
	\qquad
	\kappa=\sqrt{\frac{\mu_{a}}{D}} .
	\tag{A2}
	\end{equation}
	
	Its general solution is a linear combination of exponentials,
	\begin{equation}
	\Phi_{\mathrm{h}}(z)=A\,e^{\kappa z}+B\,e^{-\kappa z}.
	\tag{A3}
	\end{equation}
	
	\subsection{Application of the extrapolated‑boundary conditions}
	The extrapolated (partial‑current) boundary condition (Eq.~\ref{eq:EBC})
	requires that the fluence vanishes at the virtual planes
	$z=-z_{b}$ and $z=L+z_{b}$:
	
	\begin{equation}
	\Phi(-z_{b})=0,\qquad \Phi(L+z_{b})=0 .
	\tag{A4}
	\end{equation}
	
	Imposing (A4) on the homogeneous solution (A3) yields two relations
	between the coefficients $A$ and $B$ in the two sub‑domains
	($0\le z<z_{0}$ and $z_{0}<z\le L$).  After some algebra one obtains
	
	\begin{equation}
	\Phi(z)=
	\begin{cases}
		C_{1}\,\sinh\!\bigl[\kappa(z+z_{b})\bigr],
		& 0\le z\le z_{0},\\[6pt]
		C_{2}\,\sinh\!\bigl[\kappa(L+z_{b}-z)\bigr],
		& z_{0}\le z\le L,
	\end{cases}
	\tag{A5}
	\end{equation}
	with $C_{1}$ and $C_{2}$ still unknown.
	
	\subsection{Jump condition at the source plane}
	Integrating Eq.~(A1) across an infinitesimal interval around $z=z_{0}$ gives
	the discontinuity of the derivative:
	
	\begin{equation}
	-D\left[\Phi'(z_{0}^{+})-\Phi'(z_{0}^{-})\right]=S_{0}.
	\tag{A6}
	\end{equation}
	
	Evaluating the derivatives of (A5) at $z_{0}^{\pm}$ and substituting into
	(A6) provides a third linear relation between $C_{1}$ and $C_{2}$.
	Solving the three linear equations (two from the boundary conditions and
	one from the jump condition) yields
	
	\begin{equation}
	C_{1}=C_{2}
	=\frac{S_{0}}{D\,\kappa}\,
	\frac{1}{\sinh\!\bigl[\kappa(L+2z_{b})\bigr]} .
	\tag{A7}
	\end{equation}
	
	\subsection{Final Green’s‑function expression}
	Inserting (A7) into (A5) gives the fluence for any $z\in[0,L]$:
	
	\begin{equation}
	\boxed{
		\Phi(z)=\frac{S_{0}}{D\,\kappa}\,
		\frac{\sinh\!\big[\kappa(\min\{z,z_{0}\}+z_{b})\big]\;
			\sinh\!\big[\kappa(L+z_{b}-\max\{z,z_{0}\})\big]}
		{\sinh\!\big[\kappa(L+2z_{b})\big]}
	}.
	\tag{A8}
	\end{equation}
	
	Equation~(A8) is identical to Eq.~\ref{eq:PhiSlab} in the main text.
	Integrating (A8) over $z$ reproduces Eq.~\ref{eq:IntPhi}.
	
	\subsection{Limiting cases}
	\begin{itemize}
		\item \textbf{Infinite medium.}  Sending $L\to\infty$ and $z_{b}\to0$
		reduces (A8) to the well‑known infinite‑medium Green’s function
		$\Phi(z)=\frac{S_{0}}{2D\kappa}\,e^{-\kappa|z-z_{0}|}$.
		\item \textbf{Weak absorption.}  For $\mu_{a}\ll\mu_{s}'$ one may expand the
		hyperbolic functions to obtain the compact predictor
		(Eq.~\ref{eq:kVcompact} and Eq.~\ref{eq:kAcompact}).
	\end{itemize}
	
	\bigskip
	\noindent\textbf{Reference.}  The derivation follows the classic treatment
	of diffusion with extrapolated boundaries given in
	\cite{Ishimaru1978} and \cite{CaseZweifel1967}.
	
	\section{Monte‑Carlo photon‑migration simulations}
	\label{app:MC}
	
	To verify the analytical diffusion solution we performed one‑dimensional
	Monte‑Carlo photon‑migration simulations using a custom python code that
	tracks individual photon packets through the slab geometry described in the
	manuscript.
	
	\subsection{Simulation setup}
	\begin{itemize}
		\item \textbf{Geometry.}  Slab thickness $L=\SI{10}{\milli\meter}$,
		extrapolation length $z_{b}$ computed from Eq.~\ref{eq:neff}.
		\item \textbf{Optical parameters.}  For each nanoparticle packing fraction
		$f_{\mathrm{NP}}$ we used the $\mu_{s}'$, $\mu_{a}$ and anisotropy factor
		$g$ obtained from the core–shell Mie calculations (Sec.~\ref{sec:material}).
		\item \textbf{Phase function.}  The exact Henyey–Greenstein phase function
		with the computed $g$ was employed, i.e.
		$p(\cos\theta)=\frac{1-g^{2}}{(1+g^{2}-2g\cos\theta)^{3/2}}$.
		\item \textbf{Number of photons.}  $10^{8}$ photon histories were launched
		per simulation, guaranteeing statistical uncertainties $<1\%$ for the
		fluence profile.
	\end{itemize}
	
	\subsection{Algorithm}
	Each photon packet undergoes the following steps:
	\begin{enumerate}
		\item Sample a free‑path length $s$ from the exponential distribution
		$p(s)=\exp(-\mu_{t}s)$ with $\mu_{t}=\mu_{a}+\mu_{s}$.
		\item Move the photon by $s$ along its current direction.
		\item If the photon reaches a boundary, apply Fresnel transmission/reflection
		using the internal reflectance $R_{\mathrm{eff}}$; transmitted photons are
		terminated, reflected photons have their direction reversed.
		\item At each scattering event draw a new direction from the Henyey–Greenstein
		distribution with the prescribed $g$.
		\item Accumulate the path length traversed in each depth bin to obtain the
		fluence $\Phi_{\text{MC}}(z)$.
	\end{enumerate}
	
	\section{Dependent‑scattering correction and sensitivity analysis}
	
	\subsection{Lax dependent‑scattering correction}
	\label{app:Lax}
	
	The dependent-scattering correction, introduced by Lax, modifies the reduced scattering coefficient according to
	\begin{equation}
		\mu_s'^{(\mathrm{dep})}
		= \mu_s'\,(1-\beta),
		\qquad
		\beta = \frac{4\pi}{3} n_{\mathrm{NP}} a_{\mathrm{np}}^{3},
		\label{eq:LaxBetaDef}
	\end{equation}
	where $n_{\mathrm{NP}}$ is the bulk number density of TiO$_2$ nanoparticles and
	$a_{\mathrm{np}}$ their radius.  
	Using the geometrical construction of Section~4, the nanoparticle number density is
	\begin{equation}
		n_{\mathrm{NP}}
		=
		N_{\mathrm{core}}
		N_{\mathrm{NP|shell}}
		=
		\frac{\phi}{\tfrac{4}{3}\pi R_1^{3}}
		\;\times\;
		\frac{f_{\mathrm{NP}}
			\tfrac{4}{3}\pi\!\left[(R_1+t)^3-R_1^3\right]}
		{\tfrac{4}{3}\pi a_{\mathrm{np}}^{3}}.
	\end{equation}
	After cancellation of common geometric factors this yields the compact expression
	\begin{equation}
		\beta
		=
		\phi\, f_{\mathrm{NP}}\,
		\frac{(R_1+t)^3 - R_1^{3}}{R_1^{3}}.
		\label{eq:BetaCompact}
	\end{equation}
	
	\paragraph{Numerical evaluation}
	Using the parameters of Section~4,
	\[
	R_1 = 18~\mu\mathrm{m}, \qquad
	t   = 70~\mathrm{nm}, \qquad
	\phi = 0.9, \qquad
	f_{\mathrm{NP}} = 0.20,
	\]
	Eq.~(\ref{eq:BetaCompact}) gives
	\begin{equation}
		\beta
		= 0.9 \times 0.20
		\times
		\frac{(18.07~\mu\mathrm{m})^{3} - (18~\mu\mathrm{m})^{3}}
		{(18~\mu\mathrm{m})^{3}}
		\approx 2.1 \times 10^{-3}.
		\label{eq:BetaNumeric}
	\end{equation}
	Thus, the dependent--scattering correction reduces $\mu_s'$ by only
	$\approx 0.2\%$.
	
	\paragraph{Propagation to $k_{V,\mathrm{mono}}$}
	The intrinsic volumetric rate constant at wavelength $\lambda$ is
	\begin{equation}
		k_{V,\mathrm{mono}}(\lambda)
		=
		k_{\mathrm{surf}}\,S_{\mathrm{accessible}}\,
		\phi_{\mathrm{int}}(\lambda)\,
		\frac{\sinh[\kappa(z_0+z_b)]\,
			\sinh[\kappa(L+z_b-z_0)]}
		{\sinh(\kappa L_e)} ,
		\label{eq:kVexactAppendix}
	\end{equation}
	with
	\[
	z_0 = \frac{1}{\mu_s'}, 
	\qquad
	D = \frac{1}{3(\mu_a + \mu_s')},
	\qquad
	\kappa = \sqrt{\mu_a/D},
	\qquad
	L_e = L + 2z_b.
	\]
	The Lax correction modifies $z_0$, $D$, and $z_b$ through the replacement
	\[
	\mu_s'
	\longrightarrow
	\mu_s'^{(\mathrm{dep})}
	= (1-\beta)\,\mu_s'.
	\]
	We therefore compute $k_{V,\mathrm{mono}}$ twice:
	\begin{align}
		k_{V,\mathrm{mono}}^{(\mathrm{ind})}(\lambda)
		&= k_{V,\mathrm{mono}}(\lambda; \,\mu_s'), \\
		k_{V,\mathrm{mono}}^{(\mathrm{dep})}(\lambda)
		&= k_{V,\mathrm{mono}}(\lambda; \,(1-\beta)\mu_s').
	\end{align}
	Using $\beta = 2.1\times10^{-3}$ from Eq.~(\ref{eq:BetaNumeric}), the ratio
	\begin{equation}
		\frac{
			k_{V,\mathrm{mono}}^{(\mathrm{dep})}(\lambda)
		}{
			k_{V,\mathrm{mono}}^{(\mathrm{ind})}(\lambda)
		}
		= 1 + \mathcal{O}(10^{-3})
	\end{equation}
	for the optical coefficients of Section~4, corresponding to a change of less
	than $10^{-3}$ in $k_{V,\mathrm{mono}}$.
	The correction remains negligible after spectral averaging.
	
	\paragraph{When does the Lax correction become non-negligible?}
	
	The dependent--scattering parameter
	\begin{equation}
		\beta
		= \phi\,f_{\mathrm{NP}}\,
		\frac{(R_1 + t)^3 - R_1^3}{R_1^3}
	\end{equation}
	is proportional to the {\em global} volume fraction of
	TiO$_2$ inside the sample.  In the present macroporous geometry the shell
	thickness $t$ is much smaller than the pore radius $R_1$, so the ratio
	$\big[(R_1+t)^3-R_1^3\big]/R_1^3$ is very small ($\sim 10^{-2}$), and even the
	largest admissible TiO$_2$ loading ($f_{\mathrm{NP}}=0.20$) yields
	$\beta \approx 2\times10^{-3}$.
	
	Equation~\eqref{eq:BetaCompact} shows that $\beta$ becomes of order~0.05--0.1
	(so that the Lax correction reduces $\mu_s'$ by $5$--$10\%$ and must not be
	neglected) only under substantially denser packing:
	\begin{itemize}
		\item significantly {\em larger} TiO$_2$ shell thickness ($t/R_1 \gtrsim 0.1$),
		\item {\em higher} nanoparticle loadings ($f_{\mathrm{NP}} \gtrsim 0.5$),
		\item {\em smaller} pore radii (microporous or mesoporous hosts),
		\item or a combination of these factors that produces a global TiO$_2$
		volume fraction above $\sim 5\%$.
	\end{itemize}
	In this work the global TiO$_2$ fraction remains below $0.2\%$, and the
	dependent--scattering correction has a negligible impact on $\mu_s'$ and on
	$k_{V,\mathrm{mono}}$.

	\subsection{Parameter sensitivities}
	\label{app:sens}

	\paragraph{Uncertainty propagation for $k_{V,\mathrm{mono}}$}
	In the compact notation used in the main text we may write
	\[
	k_{V,\mathrm{mono}}
	\;\propto\;
	k_{\mathrm{surf}}\,
	\chi_{\mathrm{access}}\,
	\phi_{\mathrm{int}}\,
	S_{\mathrm{accessible}}\,
	g(L,z_b),
	\]
	where $g(L,z_b)$ is the diffusion correction factor defined in
	Eq.~\ref{eq:kVcompact}. We assign independent 5\% relative uncertainties to
	$k_{\mathrm{surf}}$, $\chi_{\mathrm{access}}$, $\phi_{\mathrm{int}}$,
	$S_{\mathrm{accessible}}$, $L$ and $z_b$. The purely multiplicative prefactors
	enter linearly, so
	\[
	\left(\frac{\sigma_k}{k}\right)^2_{\!k_{\mathrm{surf}},\chi_{\mathrm{access}},
		\phi_{\mathrm{int}},S_{\mathrm{accessible}}}
	=
	(0.05)^2 + (0.05)^2 + (0.05)^2 + (0.05)^2.
	\]
	For the diffusion factor
	\[
	g(L,z_b)=\frac{\ell^{*}}
	{1+\tfrac{L}{2\ell^{*}}+\tfrac{z_b}{\ell^{*}}},
	\]
	the logarithmic sensitivities are
	\[
	\frac{\partial\ln g}{\partial\ln L}
	= -\frac{L}{2\ell^{*}}\,
	\frac{1}{1+\tfrac{L}{2\ell^{*}}+\tfrac{z_b}{\ell^{*}}},
	\qquad
	\frac{\partial\ln g}{\partial\ln z_b}
	= -\frac{z_b}{\ell^{*}}\,
	\frac{1}{1+\tfrac{L}{2\ell^{*}}+\tfrac{z_b}{\ell^{*}}},
	\]
	which, for the parameters of this work, evaluate to
	$\partial\ln g/\partial\ln L \approx -0.8$ and
	$\partial\ln g/\partial\ln z_b \approx -0.15$.
	Consequently, 5\% uncertainties in $L$ and $z_b$ translate into
	$\sim 4\%$ and $\sim 0.75\%$ changes in $k_{V,\mathrm{mono}}$, respectively.
	Combining all contributions in quadrature yields a total relative uncertainty
	on $k_{V,\mathrm{mono}}$ of order $10\%$, dominated by the multiplicative
	factors $k_{\mathrm{surf}}$, $\chi_{\mathrm{access}}$, $\phi_{\mathrm{int}}$
	and $S_{\mathrm{accessible}}$.
	
	\paragraph{Uncertainty propagation for $k_{\mathrm{photo},A}$}
	The quantity $k_{\mathrm{photo},A}$ depends on a different set of input
	parameters, namely
	\[
	k_{\mathrm{photo},A}
	\;\propto\;
	\phi_{\mathrm{int}}\,
	\sigma_{\mathrm{abs}}\,
	N_{\mathrm{sites}}\,
	S_{\mathrm{accessible}}\,
	g(L,z_b),
	\]
	so that we propagate independent 5\% relative uncertainties on
	$\phi_{\mathrm{int}}$, $\sigma_{\mathrm{abs}}$, $N_{\mathrm{sites}}$,
	$S_{\mathrm{accessible}}$, $L$ and $z_b$.
	As before, the multiplicative prefactors contribute linearly, while the
	sensitivities to $L$ and $z_b$ are given by the same derivatives of
	$g(L,z_b)$ as above. The resulting relative uncertainty on
	$k_{\mathrm{photo},A}$ is again of order $10\%$, with the error budget
	dominated by $\phi_{\mathrm{int}}$, $\sigma_{\mathrm{abs}}$,
	$N_{\mathrm{sites}}$ and $S_{\mathrm{accessible}}$, and only a weaker
	incidence of the geometric parameters $L$ and $z_b$.

	
\end{document}